\documentclass[runningheads]{llncs}
\usepackage[T1]{fontenc}
\usepackage{graphicx}
\usepackage{amssymb, amsmath}
\usepackage{booktabs}
\usepackage{multirow}
\usepackage{graphicx}
\usepackage[acronym]{glossaries}
\usepackage{hyperref}
\usepackage{array} % For column formatting
\usepackage{subcaption}
\usepackage{siunitx}
\usepackage{graphicx}
\usepackage{xcolor}
\usepackage{cleveref}

\usepackage{color}

\begin{document}
%

% Acronyms

\newacronym{MSE}{MSE}{Mean Squared Error}
\newacronym{GV}{GV}{Gradient Variance}
\newacronym{GEE}{GEE}{Gaussian Edge-Enhanced}
\newacronym{CT}{CT}{Computed Tomography}
\newacronym{TV}{TV}{Total Variation}
\newacronym{ART}{ART}{Algebraic Reconstruction Technique}
\newacronym{DFT}{DFT}{Discrete Fourier Transform}
\newacronym{FBP}{FBP}{Filtered Backprojection}
\newacronym{SSIM}{SSIM}{Structural Similarity Index Measure}
\newacronym{PSNR}{PSNR}{Peak Signal-to-Noise Ratio}
\newacronym{FOV}{FOV}{Field-of-View}
\newacronym{CBCT}{CBCT}{Cone-Beam Computed Tomography}
\newacronym{FDK}{FDK}{Feldkamp-Davis-Kress Algorithm}

\title{EAGLE: An Edge-Aware Gradient Localization Enhanced Loss for CT Image Reconstruction}
\titlerunning{EAGLE: An Edge-Aware Gradient Localization Enhanced Loss}
\author{Yipeng Sun\inst{1} \and
Yixing Huang\inst{2} \and
Linda-Sophie Schneider\inst{1} \and
Mareike Thies\inst{1} \and
Mingxuan Gu\inst{1} \and
Siyuan Mei\inst{1} \and
Siming Bayer\inst{1} \and
Andreas Maier\inst{1}
}
\authorrunning{Sun et al.}
\institute{Friedrich-Alexander-Universität Erlangen-Nürnberg, Erlangen, Germany \and
University Hospital Erlangen, Erlangen, Germany\\
\email{yipeng.sun@fau.de}
}
\maketitle              % typeset the header of the contribution
\begin{abstract}
Computed Tomography (CT) image reconstruction is crucial for accurate diagnosis and deep learning approaches have demonstrated significant potential in improving reconstruction quality. However, the choice of loss function profoundly affects the reconstructed images. Traditional mean squared error loss often produces blurry images lacking fine details, while alternatives designed to improve may introduce structural artifacts or other undesirable effects. To address these limitations, we propose Eagle-Loss, a novel loss function designed to enhance the visual quality of CT image reconstructions. Eagle-Loss applies spectral analysis of localized features within gradient changes to enhance sharpness and well-defined edges. We evaluated Eagle-Loss on two public datasets across low-dose CT reconstruction and CT field-of-view extension tasks. Our results show that Eagle-Loss consistently improves the visual quality of reconstructed images, surpassing state-of-the-art methods across various network architectures. Code and data are available at \url{https://github.com/sypsyp97/Eagle_Loss}.

\end{abstract}
\section{Introduction}
\gls{CT} images hold significant importance in modern healthcare, contributing to the diagnosis and treatment of various diseases. Their reconstruction quality and speed are essential for patient care and clinical efficiency. Traditionally,  analytical methods have been employed for reconstruction, but they can fall short in speed or image quality. To address this, there has been a surge in the leveraging of deep learning to improve reconstruction in various aspects \cite{maier2019gentle,wang2024sub2full,chen2023self,wang2020deep,yu2019ea,weimin2024enhancing,chi2023low,pan2024reconstruction}. Deep learning approaches achieve optimization by minimizing a specific objective function, known as the loss function, through backpropagation. Therefore, the careful design of the loss function is critical for the reconstructed \gls{CT} images.

Pixel-wise \gls{MSE} is commonly used in \gls{CT} image reconstruction due to its straightforward computation and alignment with Gaussian noise models. However, \gls{MSE} may not accurately reflect human-perceived image quality \cite{zhang2012comprehensive}. This shortfall arises because \gls{MSE} ignores how noise perception varies with image content, such as luminance and contrast \cite{wang2004image,zhao2016loss}, to which the human visual system is sensitive. 

For countering these shortcomings of \gls{MSE}, various alternative loss functions have been proposed \cite{stimpel2019projection,abrahamyan2022gradient,johnson2016perceptual,seif2018edge,jiang2021focal,ge2023g,ledig2017photo,benjdiraa2023guided,fu2022edge,wang2020depth}, which can be categorized into three main types: perceptual-driven loss functions, gradient-based methods and approaches focus on frequency domain. Perceptual loss, which leverages pre-trained neural networks for feature comparison, can bias reconstructed images towards the style of its training data \cite{ma2020structure}. This is particularly problematic for \gls{CT} images, where stylistic deviations can obscure important details. While transfer learning (pre-training the perceptual loss model on \gls{CT} data) can mitigate this issue \cite{han2022perceptual}, it still introduces significant computational overhead. The model must be called for every loss calculation during both training and validation, slowing down the overall process. In contrast, gradient-based loss and frequency domain methods require markedly fewer computations. However, current gradient-based techniques tend to examine gradients only in the spatial domain, which can lead to globally blurred gradient maps and hinder the reconstruction of sharp edges. Similarly, common frequency domain methods focus solely on magnitude differences between the reconstructed and ground truth images by Fourier transform \cite{sun2024data}. These approaches can recover high-frequency texture details well, but struggle to accurately define edges due to the omission of phase information.

Given the challenge of optimizing phase information concurrently with magnitude, which is crucial for achieving clear edges and image sharpness during reconstruction, this study introduces a new loss function termed "Eagle-Loss". This innovative approach enhances localization by segmenting the gradient map into non-overlapping patches. Within these patches, we compute the intra-block variance to form a novel variance map. This variance map is then analyzed in the frequency domain. To the best of our knowledge we are the first to apply frequency analysis to localized features within gradient maps.

We conducted comprehensive evaluations of Eagle-Loss on two public datasets, across two \gls{CT} reconstruction scenarios: low-dose \gls{CT} reconstruction and \gls{CT} \gls{FOV} extension. In low-dose \gls{CT}, Eagle-Loss was integrated into two deep learning models and also employed as a regularizer in \gls{ART} \cite{gordon1970algebraic}. For \gls{CT} \gls{FOV} extension, we incorporated Eagle-Loss into a generative network to assess its efficacy. Notably, a public dataset was modified for \gls{CT} \gls{FOV} extension task and made open source. Our results show that Eagle-Loss outperforms state-of-the-art loss functions in terms of visual quality and sharpness. This model-independent approach offers a reliable solution to the challenges prevalent in current \gls{CT} reconstruction methodologies.

% Paper structure
The paper is structured as follows: \Cref{sec:methodology} provides an in-depth discussion of the motivation for Eagle-Loss as well as its mathematical foundations. \Cref{sec:experiments} outlines the experimental setup. \Cref{sec:results} presents the results of our experiments and an ablation study of the hyperparameter in Eagle-Loss. Finally, \cref{sec:conclusions} summarizes the contributions of our work.

\section{Methodology}
\label{sec:methodology}

\begin{figure}[t]
    \centering
    \includegraphics[width=1.0\columnwidth]{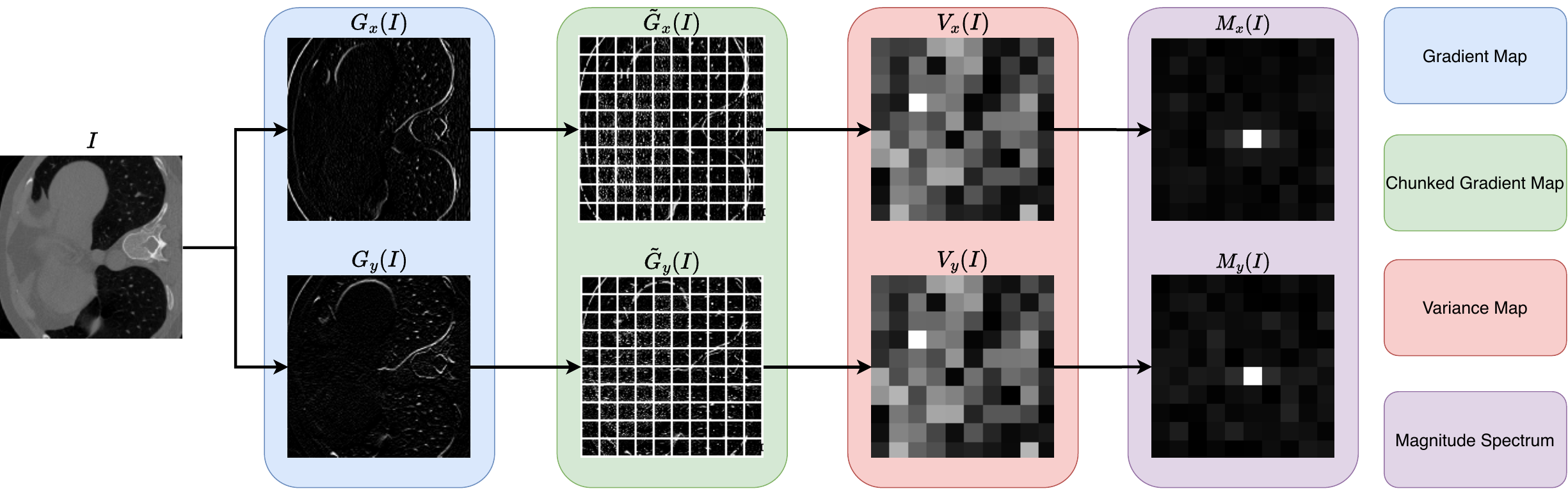}
    \caption{Illustration of steps involved in computing the magnitude spectrum \( M_x(I) \) and \( M_y(I) \) from an image \( I \). In this figure, larger patches are used for better visualization.}
    \label{fig:loss_illus}
\end{figure}

Eagle-Loss is inspired by the observation that blurring in a reconstructed image typically reduces the variance in the patches of its gradient map \cite{abrahamyan2022gradient}. This reduction results in a reconstructed image with variance maps that have less high-frequency detail compared to the variance maps of the original image. Therefore, we choose to high-pass filter the variance maps and perform a magnitude spectral analysis. The entire process of computing the magnitude spectrum is illustrated in Fig. \ref{fig:loss_illus}. Our study primarily focuses on grayscale images. Consequently, our mathematical formulations are tailored to images represented in the real number space $\mathbb{R}^{w \times h}$.

For a given image $I \in \mathbb{R}^{w \times h}$, we initiate our methodology by computing its gradient maps, which are denoted as \( G_x(I) \) and \( G_y(I) \), corresponding to the gradients along the $x$ and $y$ axes. In this work, the calculation employs Scharr kernels \( K_x \) and \( K_y \)  \cite{scharr2000optimal} to capture the edge details in both horizontal and vertical orientations. This process is defined as
\begin{align}
G_x(I) = I * K_x, \quad G_y(I) = I * K_y,
\end{align}
where \( * \) symbolizes the convolution operation. 

Following the convolution, the gradient maps are divided into non-overlapping patches, a process denoted by \(\mathcal{U}\). Each patch has the dimension of \(n \times n\). The conversion of gradient maps into their patched counterparts is expressed as
\begin{align}
\label{eq:unfold}
\tilde{G}_x(I) = \mathcal{U}\left\{G_x(I), n\right\}, \quad \tilde{G}_y(I) = \mathcal{U}\left\{G_y(I), n\right\},
\end{align}
yielding the patched gradient maps \( \tilde{G}_x(I) \) and \( \tilde{G}_y(I) \), which extract localized gradient information across the image.

For each patch $\mathcal{P}^{x}_{i,j}$ and $\mathcal{P}^{y}_{i,j}$ $\in$ $\mathbb{R}^{n \times n}$, situated at the $i$-th row and $j$-th column in \( \tilde{G}_x(I) \) and \( \tilde{G}_y(I) \), we calculate the variance maps $V_x(I)$ and $V_y(I)$ via
\begin{equation}
\begin{aligned}
v^{x}_{i,j} &= \sigma^2\left(\mathcal{P}^{x}_{i,j}\right), \quad v^{y}_{i,j} &= \sigma^2\left(\mathcal{P}^{y}_{i,j}\right), \quad  &i \in \left\{1, 2, \ldots, \frac{w}{n}\right\}, \quad j \in \left\{1, 2, \ldots, \frac{h}{n}\right\}.
\end{aligned}
\end{equation}
Here, $v^{x}_{i,j}$ and $v^{y}_{i,j}$ quantify the variances within the respective patches, with $\sigma^2$ indicating the variance calculation.

The next phase involves the application of a \gls{DFT} to $V_x(I)$ and $V_y(I)$, followed by the implementation of a Gaussian high-pass filter in the frequency domain. This process yields the magnitude spectrum $M_x(I)$ and $M_y(I)$, formalized as
\begin{equation}
\label{eq:filtering}
\begin{aligned}
M_x(I) = W \odot \left| \mathcal{F}\left\{V_x(I)\right\} \right|, \quad
M_y(I) = W \odot \left| \mathcal{F}\left\{V_y(I)\right\} \right|,
\end{aligned}
\end{equation}
where \( \mathcal{F} \) is the \gls{DFT} and \( \odot \) represents element-wise multiplication. The Gaussian high-pass filter, denoted as \( W \), is formulated by
\begin{equation}
W = 1 - e^{-\frac{(\sqrt{f_x^2 + f_y^2} - \kappa)^2}{2}},
\label{eq:high_pass filter}
\end{equation}
with \( f_x \) and \( f_y \) being the frequency components in the $x$ and $y$ directions, and \( \kappa \) setting the cutoff frequency.

For the reconstructed image $I_{rec}$ and the ground truth image $I_{g}$, our proposed Eagle-Loss is finally computed as
\begin{equation}
\mathcal{L}_{Eagle} = \frac{1}{N} \Big\lVert M_x(I_{rec}) - M_x(I_{g}) \Big\lVert_1 + \frac{1}{N} \Big\lVert M_y(I_{rec}) - M_y(I_{g}) \Big\lVert_1,
\end{equation}
where $N= \frac{wh}{n^2}$ is the number of pixels in magnitude spectrum. This loss function quantifies the difference between the reconstructed image and the ground truth by calculating the $L_1$ loss of their respective magnitude spectrum. The choice of the $L_1$ norm over the $L_2$ norm is supported by Parseval's theorem \cite{parseval1806memoire}, which indicates that the Fourier transform is unitary. This property implies that the sum of the squares of function values remains invariant even after a Fourier transform is applied.

\section{Experiments}
\label{sec:experiments}

\subsubsection{Setup}
Our experiment framework was developed based on Python 3.10 and PyTorch 2.0. For optimization, we employed the Adam optimizer ($\beta_1 = 0.9$, $\beta_2 = 0.99$), starting with an initial learning rate of $1 \times 10^{-3}$. A dynamic learning rate was implemented using a OneCycle learning rate scheduler, varying between $1 \times 10^{-3}$ and $5 \times 10^{-3}$. All models were trained over 100 epochs on an NVIDIA RTX A6000 GPU. To quantitatively assess the reconstruction performance, we chose \gls{SSIM} and \gls{PSNR} as our primary evaluation metrics.

A hybrid loss function, integrating \gls{MSE} and our novel Eagle-Loss, was employed for training the models. The loss function is formally defined as

\begin{equation}
\mathcal{L} = \mathcal{L}_{\text{MSE}} + \lambda \mathcal{L}_{\text{Eagle}},
\end{equation}
where $\lambda = 1 \times 10^{-3}$ denotes the weight coefficient for Eagle-Loss, ensuring an equitable contribution of each component to the overall model performance. Moreover, a patch size $n=3$ (referenced in Eq. \ref{eq:unfold}) was selected to augment the sensitivity of the model to high-frequency based on our empirical study.

\subsubsection{Low-dose CT Reconstruction}
We utilized the LoDoPaB-CT dataset \cite{leuschner2021lodopab} for low-dose \gls{CT} reconstruction, which comprises $35,802$ training, $3,522$ validation, and $3,553$ testing samples. This dataset offers a rich collection of $362 \times 362$ phantom images and their corresponding $1000 \times 513$ sinograms.

\begin{figure}[t]
    \centering
    \includegraphics[width=\columnwidth]{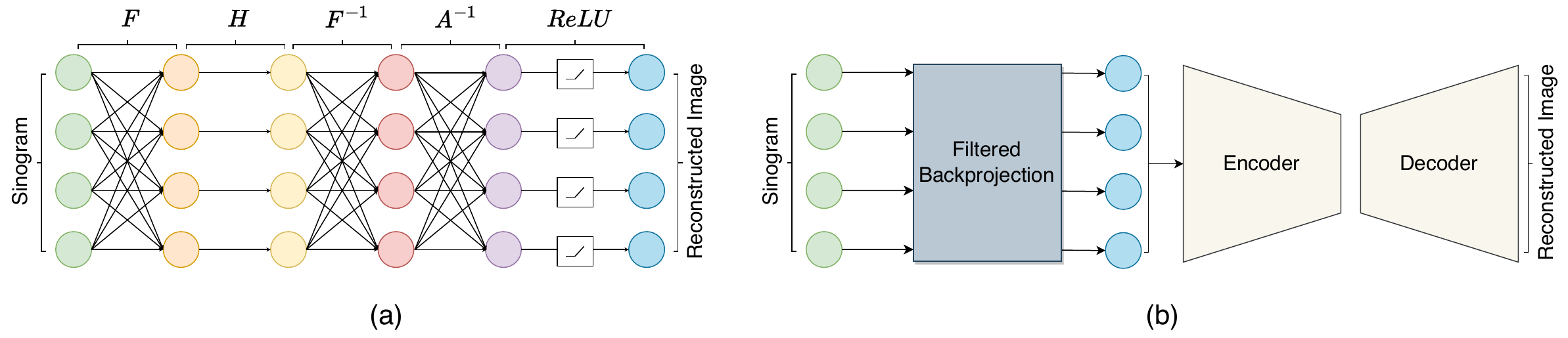}
    \caption{(a) The structure of TF-FBP. Here $H$ represents the trainable filter, \( A^{-1} \) represents the differentiable backprojection operator, while \( F^{-1} \) and \( F \) denote the inverse and forward \gls{DFT}. The differentiable backprojection operator was implemented using PYRO-NN \cite{syben2019pyro}. (b) RED-CNN with its encoder-decoder structure. The input of the encoder is the reconstructed image using \gls{FBP}.}
    \label{fig:network}
\end{figure}

Our Eagle-Loss was evaluated on two deep learning architectures: TF-FBP \cite{sun2024data} and RED-CNN \cite{chen2017low}. TF-FBP enhances the traditional \gls{FBP} algorithm with a data-driven filter, utilizing trainable coefficients of the Fourier series. In contrast, RED-CNN improves FBP-reconstructed images through an encoder-decoder framework. The network structures for both models are depicted in Fig. \ref{fig:network}. Additionally, we explored the validity of Eagle-Loss as a regularizer in \gls{ART} reconstructions. 

\subsubsection{\gls{CT} \gls{FOV} Extension}
The SMIR dataset \cite{kistler2013virtual} containing head and neck \gls{CT} data from $53$ patients was used for \gls{CT} \gls{FOV} extension \cite{huang2021data}. We simulated a \gls{CBCT} system with an \gls{FOV} size of \SI{32}{\centi\meter} to generate \gls{CBCT} projection. The 3D \gls{FDK} reconstruction with water cylinder extrapolation was used to compute the reconstruction from truncated projection data. The reconstruction volumes have a size of $512 \times 512 \times 512$ with a voxel size of $1.27\,\textrm{mm} \times 1.27\,\textrm{mm} \times 1.27\,\textrm{mm}$. The original \gls{FOV} diameter is \SI{32}{\centi\meter} and a U-Net \cite{ronneberger2015u} was used to restore missing anatomical structures with a large \gls{FOV} diameter of \SI{65}{\centi\meter}. We use $2,651$ 2D slices from $51$ patients for training, $52$ slices from one patient for validation, and $52$ slices from one patient for testing.

\section{Results}
\label{sec:results}
\subsubsection{Low-dose \gls{CT} Reconstruction}

\begin{figure}[t]
    \centering
    \begin{subfigure}{1.0\columnwidth}
        \centering
        \includegraphics[width=\textwidth]{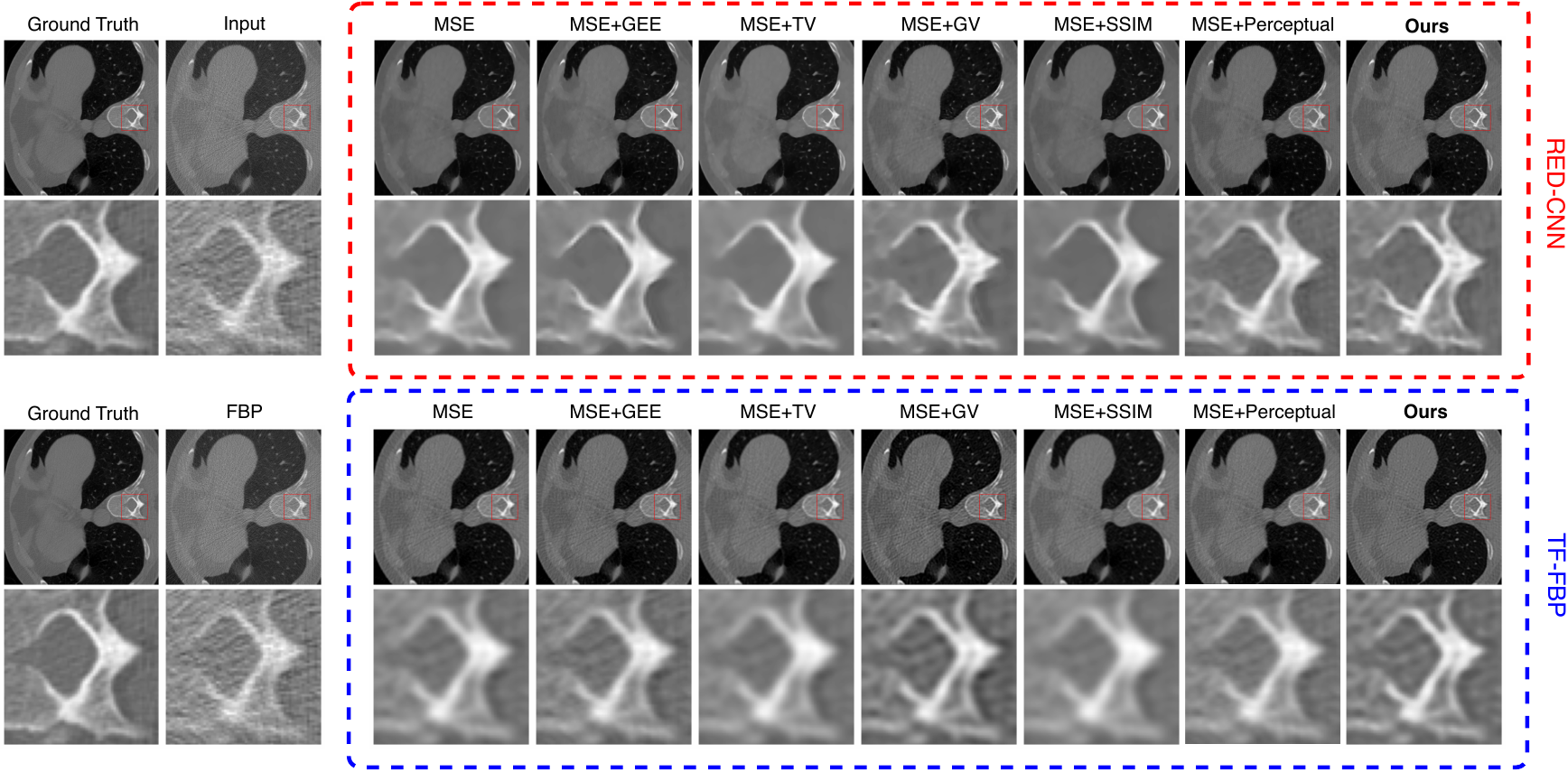}
        \caption{}
        \label{fig:ldct_sub1}
    \end{subfigure}

    \begin{subfigure}{1.0\columnwidth}
        \centering
        \includegraphics[width=\textwidth]{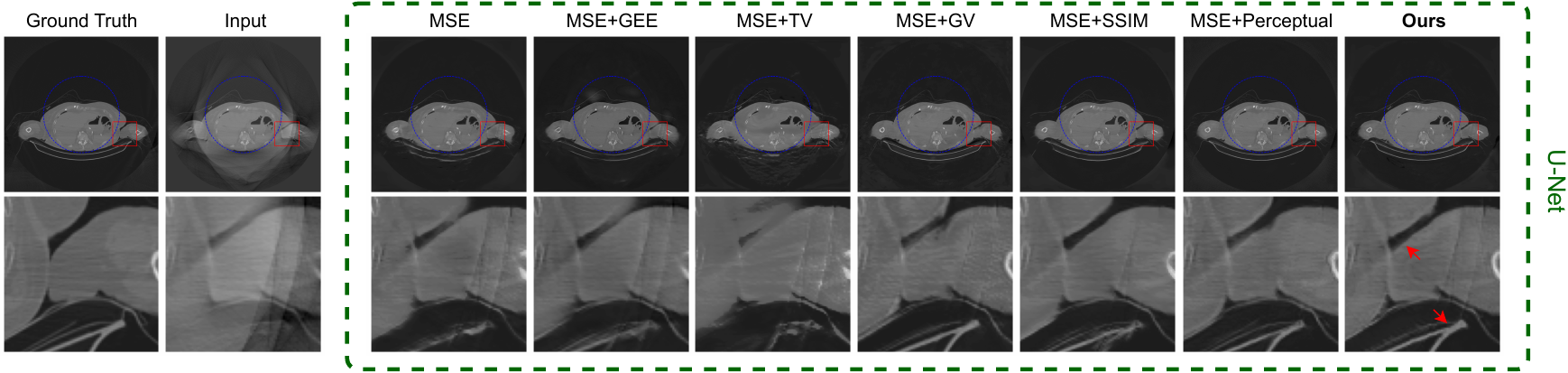}
        \caption{}
        \label{fig:fov_sub2}
    \end{subfigure}
    \caption{Comparison of our method with \gls{MSE}, \gls{GEE} \cite{sun2024data}, \gls{TV} \cite{gatys2016image}, \gls{GV} \cite{abrahamyan2022gradient}, \gls{SSIM} and Perceptual Loss \cite{johnson2016perceptual}. (a) Visualization results for low-dose CT reconstruction on LoDoPaB-CT dataset. The Red and blue dashed boxes present RED-CNN and TF-FBP. Input image of RED-CNN is the reconstruction from \gls{FBP} algorithm with a Ramp filter. (b) Visualization of \gls{CT} \gls{FOV} extension results on SMIR dataset. The blue dotted circle denotes the original \gls{FOV} boundary.}
    \label{fig:combined_figures}
\end{figure}
In our evaluation of various loss functions for low-dose \gls{CT} reconstruction, as detailed in Table \ref{tab:ldct_results}, we recognized notable variations in performance across different metrics. The \gls{MSE} loss set a robust baseline, particularly in \gls{PSNR} for the TF-FBP model. Crucially, the integration of \gls{MSE} with our novel Eagle-Loss significantly outperformed other configurations in terms of \gls{SSIM}, recording the highest values for both TF-FBP and RED-CNN. This underscores the exceptional proficiency of Eagle-Loss in preserving structural integrity. It is important to note that while Eagle-Loss did not yield the highest \gls{PSNR} scores, this is attributed to its sensitivity to high-frequency details. In contrast, other loss functions tend to over-smooth the image, resulting in artificially inflated \gls{PSNR} values. The Eagle-Loss approach prioritizes detail preservation over smoothing. This is essential in \gls{CT} imaging, where accurate detail is crucial for diagnosis. Visual comparisons of different loss function results are shown in Fig. \ref{fig:ldct_sub1}.

\begin{table}[t]
\centering
\begin{minipage}{0.5\textwidth}
\caption{Low-Dose CT Reconstruction: Performance of Loss Functions in SSIM and PSNR across different models.}
\label{tab:ldct_results}
\resizebox{!}{1.35cm}{ 
\begin{tabular}{@{}lcccccc@{}}
\toprule
\multirow{2}{*}{\textbf{Loss Function}} & \multicolumn{2}{c}{\textbf{SSIM $\uparrow$}} & \multicolumn{2}{c}{\textbf{PSNR (dB) $\uparrow$}} \\
& \text{TF-FBP} & \text{RED-CNN} & \text{TF-FBP} & \text{RED-CNN} \\ 
\midrule
MSE & 0.9558 & 0.9560 & \textbf{34.8599} & 36.4941 \\
MSE + GEE \cite{sun2024data} & 0.9526 & 0.9453 & 34.6897 & 36.6589 \\
MSE + \gls{TV} \cite{gatys2016image} & 0.9541 & 0.9709 & 34.7812 & \textbf{37.2962} \\
MSE + GV \cite{abrahamyan2022gradient} & 0.9518 & 0.9521 & 33.1900 & 35.5577 \\
MSE + SSIM& 0.9455 & 0.9659 & 34.2453 & 36.1995 \\
MSE + Perceptual \cite{johnson2016perceptual} & 0.9533 & 0.9613 & 34.4894 & 36.6393\\
\textbf{MSE + Eagle} & \textbf{0.9581} & \textbf{0.9719} & 33.6211 & 36.2331 \\
\bottomrule
\end{tabular}}
\end{minipage}%
\hfill
\begin{minipage}{0.39\textwidth}
\caption{CT FOV Extension: Performance of Loss Functions with in SSIM and PSNR.}
\label{tab:fov_results}
\resizebox{!}{1.35cm}{ 
\begin{tabular}{@{}lccc@{}}
\toprule
\textbf{Loss Function} & \textbf{SSIM $\uparrow$} & \textbf{PSNR (dB) $\uparrow$} \\
\midrule
MSE & 0.9552 & 29.1742 \\
MSE + \gls{GEE} & 0.9342 & 27.0759 \\
MSE + \gls{TV} & 0.9283 & 27.6538 \\
MSE + \gls{GV} & 0.9546 & 28.9472 \\
MSE + \gls{SSIM} & 0.9639 & 30.2405 \\
MSE + Perceptual & 0.9509 & 28.9002 \\
\textbf{MSE + Eagle} & \textbf{0.9656} & \textbf{30.9639} \\
\bottomrule
\end{tabular}}
\end{minipage}
\end{table}

\begin{figure}[t]
    \centering
    \includegraphics[width=1.0\columnwidth]{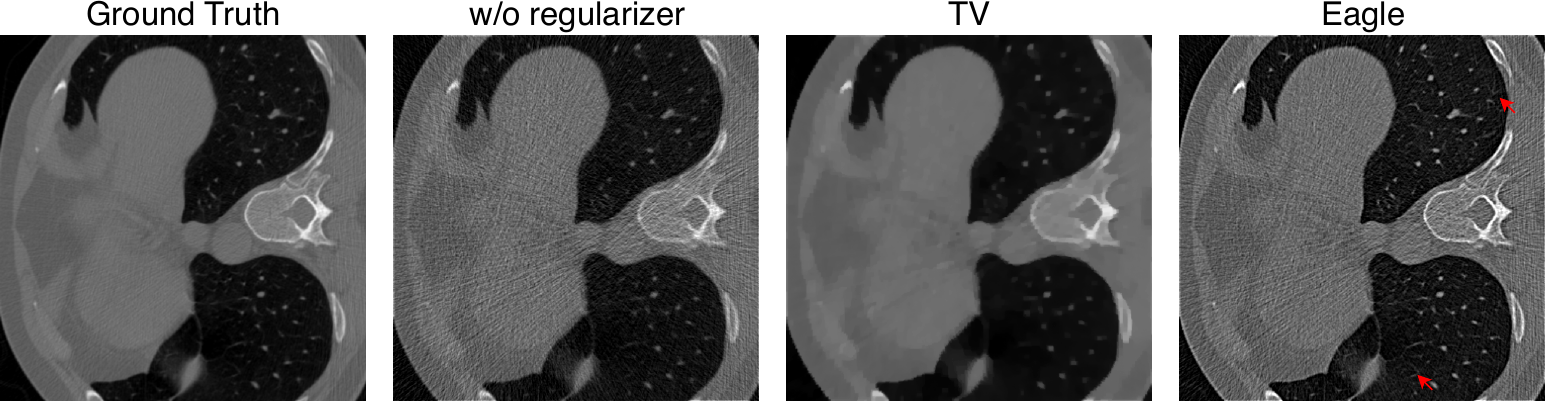}
    \caption{Comparison of Eagle-Loss regularization versus \gls{TV} regularization and no regularization in the \gls{ART} reconstruction on LoDoPaB-CT dataset.}
     \label{fig:art}
\end{figure}

In the field of \gls{CT} image reconstruction, \gls{ART} is a fundamental method. We also integrated the novel Eagle-Loss as a regularization term into the \gls{ART} algorithm. As shown in Fig. \ref{fig:art}, the inclusion of Eagle-Loss significantly improves the sharpness and fidelity of the reconstruction. Compared with the conventional \gls{TV} regularization, which often imparts a patchy character to the image, our Eagle-Loss demonstrates a superior capability for edge retention and definition. This capability demonstrates significant value in the reconstruction of complex anatomical structures, where the precise delineation of boundaries is essential. The ability of Eagle-Loss to preserve high-frequency information positions it as a powerful alternative to conventional regularizers in \gls{CT} reconstruction, leading to images with enhanced fidelity to the underlying anatomy.

\subsubsection{\gls{CT} \gls{FOV} Extension}

Table \ref{tab:fov_results} presents a detailed comparison of the results for our \gls{CT} \gls{FOV} extension task. In this task, Eagle-Loss demonstrates superior performance, achieving the highest \gls{SSIM} score and \gls{PSNR}. This success can be attributed to the fact that this reconstruction process involves less high-frequency noise and mainly focuses on compensating for the larger low-frequency artifacts. Fig. \ref{fig:fov_sub2} further illustrates the advantages of Eagle-Loss, showcasing clearer structural boundaries in the reconstructed images compared to those produced by other loss functions. However, it is noteworthy that Eagle-Loss can be less effective in addressing streaky artifacts. This limitation may arise from the high-pass filter in the variance map (referenced in Eq. \ref{eq:filtering}), in which the Gaussian filter tends to treat such artifacts as low-frequency features. Consequently, these features are filtered out, resulting in less compensation for streaky artifacts.

\subsubsection{Ablation Study}

This ablation study focuses on analyzing the effect of varying cutoff frequencies, denoted as $\kappa$, within the Gaussian high-pass filter (referenced in Eq. \ref{eq:high_pass filter}) on the quality of reconstructed images. We employed the TF-FBP model for this analysis due to its compact parameter space, where only $255$ parameters define the outcome. In TF-FBP, the quality of the reconstructed images is directly linked to the characteristics of the learned filter. 

\begin{figure}[t]
    \centering
    \includegraphics[width=1.0\columnwidth]{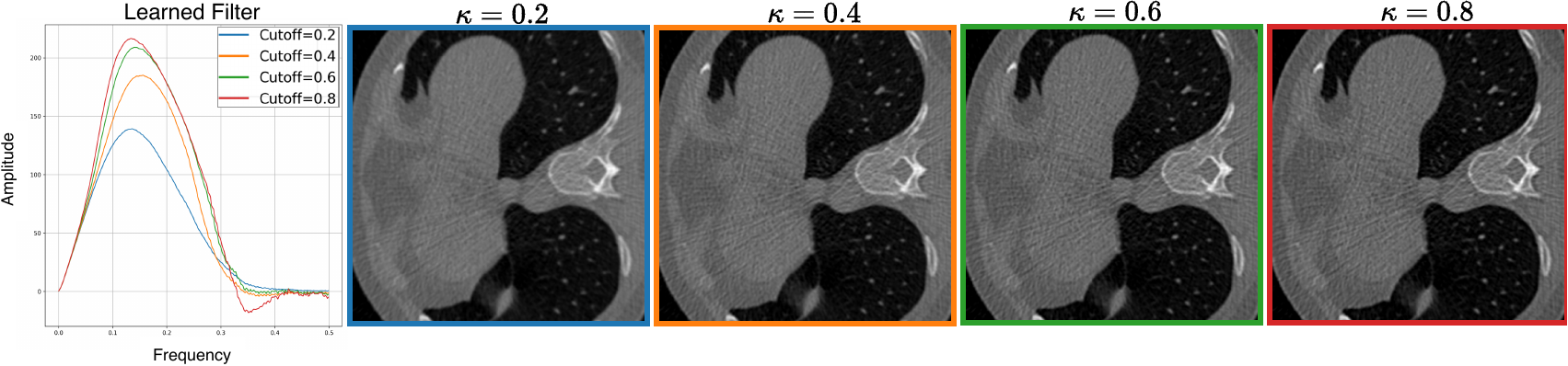}
    \caption{Visual representation of the filters and their corresponding reconstructed images at varying $\kappa$ values in the TF-FBP model. This illustration highlights the direct influence of the cutoff frequency on the enhancement of high-frequency details in image reconstruction.}
    \label{fig:ablation}
\end{figure}

Fig. \ref{fig:ablation} showcases the resultant filters and corresponding reconstructed images for different values of $\kappa$. A notable observation from our study is that higher values of $\kappa$ significantly enhance the sharpness of the reconstructed image while retaining more of the high-frequency noise.

\section{Conclusion}
\label{sec:conclusions}
In this paper, we propose Eagle-Loss, a novel loss function specifically designed to improve the quality of reconstructed images by emphasizing high-frequency details crucial for accurate edge and texture representation. Eagle-Loss leverages localized feature analysis within gradient maps and applies frequency analysis. Our experimental results demonstrate that Eagle-Loss effectively reduces image blur and enhances edge sharpness, leading to reconstructed images that exhibit superior fidelity to the ground truth. Eagle-Loss also faces limitations due to its sensitivity to hyperparameter settings. The optimal selection of these parameters varies across reconstruction tasks. Future efforts will concentrate on devising an automated hyperparameter optimization strategy, thereby enhancing the versatility and efficacy of Eagle-Loss in diverse imaging contexts. Our findings suggest that Eagle-Loss holds significant promise for \gls{CT} image reconstruction and has the potential for broader applications in other fields.

\subsubsection{Acknowledgements}
This research was financed by the ``Verbundprojekt 05D2022 - KI4D4E: Ein KI-basiertes Framework für die Visualisierung und Auswertung der massiven Datenmengen der 4D-Tomographie für Endanwender von Beamlines. Teilprojekt 5.'' (Grant number: 05D23WE1).

\bibliographystyle{splncs04}

\bibliography{mybib.bib}

\end{document}